\documentclass[preprint2]{aastex}
\usepackage{epsfig}

\shorttitle{${}^{15}\textrm{O}(\alpha,\gamma){}^{19}\textrm{Ne}$ in X-ray bursts and superbursts}
\shortauthors{Fisker, Davids, G{\"o}rres, and Wiescher}
\begin{document}
\bibliographystyle{apj}
\title{The importance of ${}^{15}\textrm{O}(\alpha,\gamma){}^{19}\textrm{Ne}$ to X-ray bursts and superbursts}
\author{Jacob Lund Fisker, Joachim G\"orres, and Michael Wiescher}
\affil{Department of Physics and Joint Institute for Nuclear Astrophysics, \\ University of Notre Dame, Notre Dame, IN 46556}
\email{jfisker@nd.edu,jgoerres@nd.edu,mwiesche@nd.edu}
\author{Barry Davids}
\affil{TRIUMF, 4004 Wesbrook Mall, Vancouver, BC V6T 2A3, Canada}
\email{davids@triumf.ca}
\begin{abstract}
One of the two breakout reactions from the hot CNO (HCNO) cycle is ${}^{15}\textrm{O}(\alpha,\gamma){}^{19}\textrm{Ne}$, which at low temperatures depends strongly on the resonance strength of the 4.033 MeV state in ${}^{19}\textrm{Ne}$. An experimental upper limit has been placed on its strength, but the lower limit on the resonance strength and thereby the astrophysical reaction rate is unconstrained experimentally. However, this breakout reaction is crucial to the thermonuclear runaway which causes type I X-ray bursts on accreting neutron stars. In this paper we exploit astronomical observations in an attempt to constrain the relevant nuclear physics and deduce a lower limit on the reaction rate. Our sensitivity study implies that if the rate were sufficiently small, accreting material would burn stably without bursts. The existence of type I X-ray bursts and superbursts consequently suggests a lower limit on the ${}^{15}\textrm{O}(\alpha,\gamma){}^{19}\textrm{Ne}$ reaction rate at low temperatures.
\end{abstract}


\section{Introduction}\label{sec:introduction}
Type I X-ray bursts were discovered independently by \cite{Grindlay76a} and \cite{Belian76} and occur in matter-transfering binary systems in which a neutron star accretes hydrogen and helium from an unevolved companion star \citep{Woosley76,Joss77,Joss78b,Taam80,Ayasli82, Lewin95, Bildsten98c}.
The transferred matter is heated to $1-2\times 10^8$ K while it slowly descends into the neutron star atmosphere as freshly infalling material continuously piles on top of it. 
At these temperatures hydrogen burns stably into helium at a constant and temperature-independent rate given by the half lives of two nuclei participating in the $\beta$-limited HCNO cycle \citep{Wallace81}.
\begin{equation}\label{eq:hcno}
\frac{dX_H}{dt}=-\frac{\lambda_{{}^{14}O}\lambda_{{}^{15}O}}{\lambda_{{}^{14}O}+\lambda_{{}^{15}O}} X_{HCNO}\,,
\end{equation}
where the decay constants $\lambda=\ln 2/T_{1/2}$ are given by the half lives of ${}^{14}\textrm{O}$ and ${}^{15}\textrm{O}$, and $X_{HCNO}$ is the total mass fraction of these two nuclei.
The triple-$\alpha$ reaction concurrently increases the mass fraction of HCNO material while depleting helium \citep{Bildsten98c}.
The mass fraction of HCNO material is important because it regulates the rate at which hydrogen burns to helium, determining the initial composition and the characteristics of the burst \citep{Fujimoto81,Fushiki87,Cumming03a,Narayan03}.

The mass fraction of HCNO material can decrease through processing into heavier isotopes via the ${}^{15}\textrm{O}(\alpha,\gamma)$${}^{19}\textrm{Ne}$ and $^{18}$Ne($\alpha,p)^{21}$Na breakout reactions \citep{Wallace81,Wiescher99}.
This can happen either during the burst, when peak temperatures exceed $10^9$ K, or during the quiescent, non-bursting phase.
At temperatures characteristic of the quiescent phase ($T \lesssim 2\times 10^8$ K) only the ${}^{15}\textrm{O}(\alpha,\gamma)$${}^{19}\textrm{Ne}$ reaction allows significant leakage out of the HCNO cycle \citep{Hahn96}. 
In this case the leak of HCNO material into heavier isotopes depends on the ratio between the $\alpha$ capture rate of ${}^{15}\textrm{O}$ and its decay rate
\begin{equation}\label{eq:leak}
R=\frac{N_A<\sigma v>\rho X_\alpha/A_\alpha}{\lambda_{{}^{15}O}}\,,
\end{equation}
where $<\sigma v>$ is the thermally averaged $^{15}$O($\alpha,\gamma)$ reaction rate per particle pair, $N_A$ is the Avogadro constant, $\rho$ is the stellar mass density, $X_\alpha$ is the ${}^4\textrm{He}$ mass fraction and $A_\alpha$ its atomic mass. 

What happens therefore depends on the relatively unknown ${}^{15}\textrm{O}(\alpha,\gamma){}^{19}\textrm{Ne}$ reaction rate: If the rate is comparable to the \cite{Caughlan88} rate, then the leak will be important in reducing the HCNO mass fraction of Eq.~[\ref{eq:hcno}] and therefore in decreasing the rate at which hydrogen burns into helium as matter sinks. With long burst recurrence times of thousands of seconds or even days \citep{Strohmayer03}, even small reaction rates can be significant, allowing hydrogen to survive to great depths. In addition, a rate of this magnitude would deplete ${}^{15}\textrm{O}$ above the ignition point during the burst runaway, ensuring that little or no ${}^{15}\textrm{O}$ survives for the next burst and requiring the accretion of a fresh layer of material.

However, the $\alpha$ width of the dominant 4.033 MeV state in ${}^{19}\textrm{Ne}$ is not well known, so the rate could be lower. If the rate is lower, there will not be any significant leak out of the HCNO cycle, whence hydrogen would burn to helium at a faster rate, depleting hydrogen from the ignition zone. 
In addition, a lower rate would prevent or weaken the breakout via ${}^{15}\textrm{O}(\alpha,\gamma)$${}^{19}\textrm{Ne}$ ensuring less energy generation during the initial stages of the burst \citep{Taam79,Ayasli82,Hanawa84,Wiescher99}. 
The survival of a significant fraction of ${}^{15}\textrm{O}$ in the upper layers would increase the hydrogen burning prior to subsequent bursts, thus altering the composition at the time of the runaway. 

In this paper we show that a sufficiently small $^{15}$O($\alpha,\gamma)$ rate can even prevent subsequent runaways, causing hydrogen and helium to burn stably. 
Stable burning creates copious amounts of ${}^{12}\textrm{C}$ and other $A\lesssim 20$ nuclei which have implications for superbursts. Astronomical observations may lead to better constraints on this rate and motivate future measurements of the ${}^{15}\textrm{O}(\alpha,\gamma){}^{19}\textrm{Ne}$ reaction rate. Such measurements would help refine current X-ray burst and superburst models.

\section{The ${}^{15}\textrm{O}(\alpha,\gamma){}^{19}\textrm{Ne}$ reaction rate}\label{sec:reaction}

For temperatures below $6\times 10^8$ K, the $^{15}\textrm{O}(\alpha,\gamma)$ reaction rate is dominated by the resonant contribution of the 3/2$^+$ state at 4.033 MeV in $^{19}$Ne, which is $\alpha$-unbound by 504 keV. Appreciable resonant contributions also come from the states at 4.379, 4.600, and 4.712 MeV. At higher temperatures contributions from higher-lying states must be included. Non-resonant capture also plays a role, but is only really important below 10$^8$ K. 

The rate was first calculated by \cite{wagoner69} and later revised by \cite{Wallace81} and \cite{Langanke86}. In this paper we provide a new estimate of the rate based on recent experiments and calculations. We calculate three rates, a best rate and upper and lower limits.

The S factor for non-resonant capture was calculated by \cite{Langanke86} within the framework of the direct capture model of \cite{Rolfs73}, adopting the known reduced $\alpha$ widths of the bound mirror states in $^{19}$F. A constant S factor of 23 MeV b was adopted by \cite{Langanke86} for the reaction rate calculation. The reactions $^{15}$N+$\alpha$ and $^{15}$O+$\alpha$ are isospin mirrors leading to analog states in $^{19}$F and $^{19}$Ne. Therefore it is justified to assume that the reduced $\alpha$ widths in $^{19}$F and $^{19}$Ne are similar, which is supported by a comparison by \cite{Davids03}. The non-resonant contribution was also calculated by \cite{dufour00} using the generator coordinate method. Despite the fact that the authors of this work state that the $E1$ transition probabilities are overestimated and that the resulting S factor should be considered an upper limit on the non-resonant contribution, they find an S factor only 85\% as large as the \cite{Langanke86} result. For the non-resonant contribution, we adopt the \cite{dufour00} calculation for our best rate, 1/3 of it for our lower limit, and use the \cite{Langanke86} calculation for our upper limit.

The contributions of resonances are proportional to $\Gamma_{\alpha}\Gamma_{\gamma}/\Gamma_{total}$ with $\Gamma_{\alpha(\gamma)}$ the $\alpha (\gamma)$ partial width and $\Gamma_{total}$ the total width of the state; for the states under consideration here, no other channels are open, so $\Gamma_{total}=\Gamma_{\alpha}+\Gamma_{\gamma}$. In addition, for the two lowest-lying states, $\Gamma_{\alpha} \ll \Gamma_{\gamma}$, and  their strengths are directly proportional to their $\alpha$ widths $\Gamma_{\alpha}$. These widths are not known for the resonances of interest and cannot be measured directly without an intense, radioactive $^{15}$O beam at low energy. 

Several indirect measurements of $\Gamma_{\alpha}$ for these resonances have been published.
\cite{Mao96} analyzed the $\alpha$ reduced width of the analog state of the 4.033 level in $^{19}$F and studied carefully the dependence of the result on the model parameters. These calculations led to an $\alpha$ width of 11.5 $\mu$eV. An alternative approach is the measurement of the $\alpha$-decay branching ratio $\Gamma_{\alpha}$/$\Gamma_{total}$. 
Two recent experiments, \cite{Davids03} and \cite{Rehm03}, found 90\% confidence level upper limits of 4.3$\times$10$^{-4}$ and 6$\times$10$^{-4}$, respectively. 
To deduce the $\alpha$ width from this ratio, the total width must also be known.
A lifetime measurement by \cite{Davidson73} and a Coulomb excitation experiment by \cite{hackman00} set 2$\sigma$ limits on the total width, 6.6 meV $\leq\Gamma_{total}\leq$ 445 meV. 
The lifetime of the mirror state in $^{19}$F corresponds to a total width of 73 meV.  
We adopt the alpha width of \cite{Mao96} for our recommended rate. 
The upper limit of 130 $\mu$eV for the $\alpha$ width is set by the measured 90\% confidence level upper limits on limits on $\Gamma_{total}$ and the $\alpha$-decay branching ratio. 
Considering the uncertainties in the determination of reduced $\alpha$ widths from states in the analog nucleus far below the Coulomb barrier, we consider an $\alpha$ width of 3\% of our recommended value, 345 neV, as a conservative lower limit.

For the $^{19}$Ne state at 4.379 MeV, \cite{Langanke86} assumed a reduced $\alpha$ width of 0.02, leading to an $\alpha$ width of 700 $\mu$eV. Three experiments report $\alpha$-decay branching ratios for this state, 0.044 $\pm$ 0.032 \citep{Magnus90}, 0.016 $\pm$ 0.05 \citep{Rehm03}, and a 90\% confidence level upper limit of 3.9$\times10^{-3}$ \citep{Davids03}. The first of these was based on poor statistics and represents an upper limit within $2\sigma$, while the energy resolution of the second measurement was not sufficient to resolve the 4.379 MeV level from nearby states, limiting the reliability of the extracted $\alpha$-decay branching ratio. Therefore we adopt the upper limit of \cite{Davids03}. The lifetime experiment of \cite{Davidson73} set a 1$\sigma$ lower limit on the total width of 5.5 meV. A  shell model calculation using the USD interaction cited in \cite{Davids03} found a radiative width of 458 meV. \cite{wilmes02} have measured the $\alpha$ width of the $^{19}$F analog state, from which \cite{Davids03} have deduced a reduced $\alpha$ width of 0.0078$^{+0.0078}_{-0.004}$. Adopting the shell model calculation for the radiative width and the reduced $\alpha$ width from the analog state leads to our recommended $\alpha$ width of 190 $\mu$eV. To calculate a lower limit on the $\alpha$ width we use the 1$\sigma$ lower limits on the total width and the reduced $\alpha$ width and find 1 $\mu$eV. Our upper limit on the $\alpha$ width is based on the shell model calculation for the radiative width and the experimental upper limit on the $\alpha$-decay branching ratio; using these values we calculate an upper limit of 2.3 meV.

\begin{figure}[thbp]
\plotone{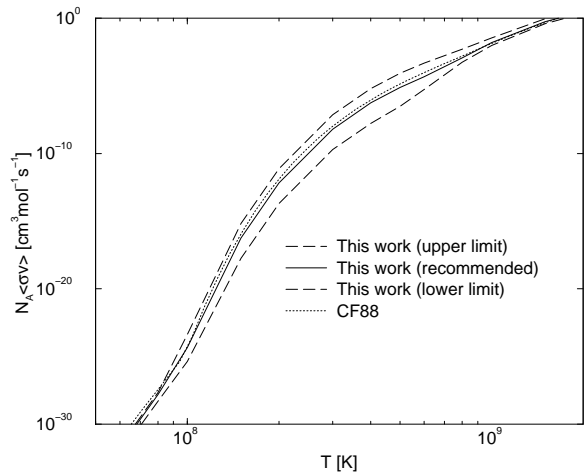}
\caption{The three ${}^{15}\textrm{O}(\alpha,\gamma)$${}^{19}\textrm{Ne}$ reaction rates adopted in this work compared to the rate of \cite{Caughlan88}.}\label{fig:o15ne19_rate}
\end{figure}

Above 1 GK, the states at 4.600, 4.712, and 5.092 MeV make appreciable contributions to the reaction rate. The parameters of these resonances were taken from \cite{Davids03}. Using these results we calculated three different rates of the $^{15}$O($\alpha,\gamma$)$^{19}$Ne reaction: a recommended rate, an upper limit and a lower limit.
The rates are shown in Fig.~\ref{fig:o15ne19_rate}. 
Model 1 represents the upper limit for the reaction rate, model 2 represents the recommended reaction rate, and model 3 represents a conservative lower limit.

\section{Computational model calculations}\label{sec:burst}
\begin{figure}[htp]
\plotone{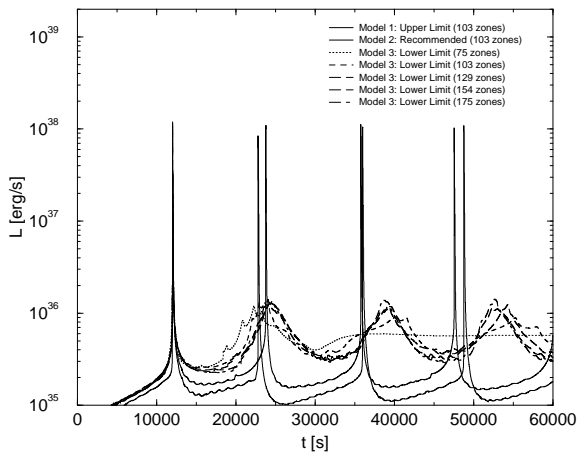}
\caption{Luminosity (as seen from an observer at infinity) as a function of (observer) time for the various models. Model 3 (lower limit) has been calculated at two different resolutions which both become stable after the first burst.}\label{fig:Lt}
\end{figure}

We calculate three X-ray burst models (Fig.~[\ref{fig:Lt}]) that differ only in their different ${}^{15}\textrm{O}(\alpha,\gamma)$${}^{19}\textrm{Ne}$ reaction rates.
These models are calculated using a new general relativistic type I X-ray burst simulation code that couples the general relativistic hydrodynamics code, \verb+AGILE+ \citep{Liebendoerfer02}, with the generic nuclear reaction network of \cite{Hix99} using the operator-split method of \cite{Henyey59}. Specifically, the calculation of a single time-step splits into four steps: 1) The nuclear reaction network is evolved in all zones assuming fixed density and temperature. 2) The change in nuclear binding energy enters \verb+AGILE+ as a source term and the hydrodynamics is evolved. This step includes the self-consistent accretion of the rest mass. This step also calculates the turbulent velocity that determines the convective mixing of the composition 3) The convective mixing of the composition is evolved by solving a simple diffusion equation based on the turbulent velocity and a mixing lenght\footnote{Since convection is extremely efficient, the exact value of $\Lambda$ is unimportant. We note that setting it equal to the pressure scale height, which is typically done in stellar evolution calculations is unphysical, since the pressure scale height of the neutron star atmosphere is much larger than the convective zone itself.}, $\Lambda=1 \textrm{cm}$ (see \cite{Weaver78}). 4) Finally, the composition is updated using the advection that was calculated in step 1 using the second-order scheme described below.

\begin{figure}[thbp]
\plotone{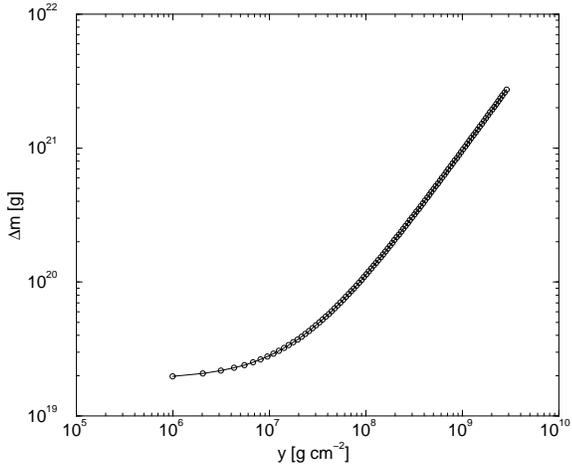}
\caption{The rest mass in a zone as a function of column density. The computational domain covers about 7 pressure scale heights and comprises 103 zones which are log-ratioed in rest mass.}\label{fig:pyzones}
\end{figure}

The code uses an arbitrarily relativistic and arbitrarily degenerate equation of state and includes radiative, conductive, and convective heat transport as described in \cite{Thorne77}. We calculate the radiative opacities due to Thomson scattering and free-free absorption using the analytic formulations of \cite{Schatz99}.
We use the same conductivity formulations for electron scattering on electrons, ions, phonons, and impurities as \cite{Brown00}. 

\verb+AGILE+ solves the general relativistic equations in a spherically symmetric geometry on a comoving grid. Fig.~[\ref{fig:pyzones}] shows how the computational domain is discretized into 103 log-ratioed grid zones with a column density\footnote{The relativistic column density is mass of a column above an area: $y\equiv\int_{R-r}^R\rho {dr \over \Gamma}$ where $\Gamma=\sqrt{1-2GM/Rc^2}$, so $P\simeq gy$, where $R$ is the neutron star radius, $M$ is the neutron star mass, $\rho$ is the density, $P$ is the pressure, and $g=GM/\Gamma R^2$ is the surface acceleration of gravity.} ranging from $y=1.2\times 10^6$ g cm$^{-2}$ to $y=3.5\times 10^9$ g cm$^{-2}$. 
In this paper, \verb+AGILE+ does not employ its adaptive grid capability. The composition is advected using a second-order total variation diminishing upwind scheme with a van Leer limiter. 

The computational domain is bounded by a realistic core boundary interface \citep{Brown03,Brown04} and a relativistically corrected grey atmosphere (see \cite{Thorne77} and \cite{Cox04}). Due to numerical constraints, previous work (e.g. \cite{Taam93,Taam96,Fisker04,Fisker04b,Fisker05b,Fisker05a,Woosley04}) has set the outer boundary at $P \sim 10^{20}\,\textrm{erg}\,\textrm{cm}^{-3}$ while assuming no nuclear burning and an opacity dominated by Thomson scattering at lower pressures.
While the assumption of neglible nuclear burning is correct, the assumption of a constant opacity is off by up to 15\%. The reason is that free-free absorption still contributes to the opacity at this pressure. 
Since the grey atmosphere model relates the opacity and the luminosity to the temperature and the temperature gradient, an inaccurate opacity at the boundary affects the temperature profile through the entire model. This approximation is perhaps acceptable at low accretion rates or very high accretion rates since it does not change the overall bursting behavior of the system, but it is very critical at the transition rate between stable burning and unstable burning as stability is determined by a very fine balance between the cooling rate of the radiation transport and the heating rate of the nuclear reactions. 
The solution is to 
use a 4th order Runge-Kutta method and numerically integrate the hydrostatic heat and pressure equations (see \cite{Thorne77} and \cite{Cox04}) from the model boundary out to $P = 10^{18}\,\textrm{erg}\,\textrm{cm}^{-3}$ where the opacity finally becomes constant.

Although heavier ashes with $A\gtrsim 100$ predominate, \emph{iff} the burst peak temperature and the hydrogen concentration at ignition are high \citep{Schatz01}, \cite{Woosley04} showed that a self-consistent X-ray burst simulation have comparably smaller peak temperatures and a smaller hydrogen concentration at the ignition point. Therefore most of the ashes concentrate around $A\approx 64$ with only a very small fraction of the nuclei reacting beyond this mass range. This means that we can use a smaller reaction network without changing the underlying burst physics of the opacity, the equation of state, or the nuclear energy release rate. 

The nuclear reaction network used in this work includes 298 isotopes (see table~[\ref{fig:sunet}]). All the connecting particle reactions are taken from REACLIB (see \cite{Sakharuk06}). 
This includes all isotopes between the valley of stability and the proton dripline up to ${}^{64}\textrm{Ge}$. Here isotopes with $\beta^+$-half lives $> 1$ day are considered ``stable'' on the timescale of the burst intervals, so their daughters are not included.
Above the ${}^{64}\textrm{Ge}$ waiting point, only isotopes between the proton drip line and half lives less than 1 minute are included. This is because protons only capture on these high-Z isotopes during the burst's peak temperature which is only sustained for a few seconds. 
Weak rates up to $Z=32$ are taken from \cite{Fuller80,Fuller82a,Fuller82b} and \cite{Langanke01}. Since only a small fraction of material is processed above $Z=32$, it is a reasonable approximation to ignore neutrino losses from heavier isotopes \citep{Schatz99}.
These considerations significantly reduce the size of the network which decreases the simulation run-time.

Since this paper investigates the transition between stable and unstable burning, we can not assume an artifical iron-atmosphere for the initial setup. The reason is that an iron-atmosphere progenitor allows more H/He to accumulate which results in a very energetic burst that could heat the atmosphere to such a degree that the thermal inertia (see \cite{Taam93}) of the initial burst would ensure that H/He would keep burning at the same rate it was accreted and thus resulting in stable burning. As iron-atmospheres never occur on accreting neutron stars in reality, the initial model was made by accreting and computing several bursts until a limit cycle equilibrium was reached. 
At this point, the composition of the burst ashes was copied to the bottom of the model-envelope after which the limit cycle equilibrium was reachieved as the model adjusted boundary conditions to the new temperature profile resulting from the changed thermal conductivity in the copied shells.
This creates a model that is physically independent of artificial starting conditions and which is approximately self-consistent with the chosen reaction library.  At the same time this avoids the need to compute the evolution long enough to replace the entire model envelope which is shown in Fig.~\ref{fig:pyzones}. 
\begin{table}[thbp]
\centering
\begin{tabular}{|r|l|r|l|r|l|}
\hline
Z & A & Z & A & Z & A\\
\hline
\hline
n & 1       & Ar& 31--38  & Kr& 69--74\\
H & 1--3    & K&  35--39  & Rb& 73--77\\  
He& 3--4    & Ca& 36--44  & Sr& 74--78\\  
Li& 7       & Sc& 39--45  &  Y& 77--82\\  
Be& 7       & Ti& 40--47  & Zr& 78--83\\  
B & 8       & V&  43--49  & Nb& 81--85\\ 
C & 12      & Cr& 44--52  & Mo& 82--86\\ 
N & 13--15  & Mn& 47--53  & Tc& 85--88\\   
O & 14--18  & Fe& 48--56  & Ru& 86--91\\
F & 17--19  & Co& 51--57  & Rh& 89--93\\ 
Ne& 18--21  & Ni& 52--62  & Pd& 90--94\\ 
Na& 20--23  & Cu& 54--63  & Ag& 94--98\\
Mg& 21--25  & Zn& 55--66  & Cd& 95--99\\
Al& 22--27  & Ga& 59--67  & In& 98--104\\
Si& 24--30  & Ge& 60--68  & Sn& 99--105\\
P&  26--31  & As& 64--69  & Sb& 106 \\ 
S&  27--34  & Se& 65--72  & Te& 107 \\  
Cl& 30--35  & Br& 68--73  & & \\
\hline
\end{tabular}
\caption{The table shows the list of isotopes which describes the $rp$-process. See the main text for details. This reaction network has been used in the following works \citep{Fisker04,Fisker04b,Fisker05b,Fisker05a}.}\label{fig:sunet}
\end{table}
We selected a global (relativistically local) accretion rate of $\dot{M}=10^{17}\textrm{g}/\textrm{s}$ ($0.09\dot{M}_{Edd}$), similar to the accretion rates used by \cite{Taam96} and \cite{Woosley04}. The accreted composition is assumed to be solar \citep{Anders89}. The underlying fiducial NS core has a gravitional mass of $1.4M_\odot$ and an areal radius of $11.2\,\textrm{km}$.

Fig.~[\ref{fig:Lt}] shows 
an initial burst for all models which can be ascribed to the fact that a limit cycle equilibrium (see \cite{Lewin95,Woosley04}) has not been reached.
The first burst of any model is usually ``special'', yet whereas models 1 and 2 both repeat this bursting behavior, model 3 turns stable after the second burst fizzles. 
As model 3 turns out to describe an oscillating form of stable helium burning in a thin shell (see \cite{Woosley04}) with a period of around 4 hours, we performed a convergence study of this model to determine the required number of zones to resolve this burning shell. Fig.~[\ref{fig:Lt}] shows that the oscillation is resolved in models having more than 100 zones in the covered range of column depths whereas 75 zones or less is insufficient. 
In the next section, we divide the models into two categories, bursting and non-bursting.

\subsection{Behavior in bursting models}
Models 1 and 2 include comparably high rates and exhibit unstable burning/bursting behavior.
During the quiescent phase the ${}^{15}\textrm{O}(\alpha,\gamma)$${}^{19}\textrm{Ne}$ reaction establishes an outflow from the HCNO cycle, thereby reducing $X_{HCNO}$ and ensuring that hydrogen burns slowly and thus survives to a great depth. 

For model 2 the thermonuclear instability criterion for a degenerate atmosphere (see e.g.~\cite{Rakavy67}) is violated at y = 7.9 $\times 10^7$ g cm$^{-2}$. 
One hundred seconds before the burst peak temperature is reached, the ratio $R$ of eq.~[\ref{eq:leak}] is $\sim 1\%$; $R$ exceeds $\sim 10\%$ about ten seconds prior to the burst which is just before the ${}^{14}\textrm{O}(\alpha,p)$${}^{17}\textrm{F}$-reaction becomes competitive and establishes the $\alpha p$-process \citep{Wallace81,Fisker04b}. After escaping the HCNO cycle the reaction flow proceeds through the heavier isotopes as described in \cite{Woosley04}.
During the runaway, the ${}^{15}\textrm{O}(\alpha,\gamma)$${}^{19}\textrm{Ne}$ reaction will eliminate most of the ${}^{15}\textrm{O}$, since ${}^{15}\textrm{O}$ is only formed via the slow ${}^{14}\textrm{O}(\beta^+, T_{1/2}=76\textrm{s})$ ${}^{14}\textrm{N}(p,\gamma)$ ${}^{15}\textrm{O}$, thus reducing $X_{HCNO}$.
This means that the conversion of hydrogen to helium for the subsequent burst will be delayed, as explained in section \ref{sec:introduction}.

The final CNO abundance would also be influenced by the amount of accreted CNO material; small values, which might arise from spallation in the atmosphere \citep{Bildsten92}, have been shown to create chaotic bursting behavior \citep{Taam93}. 

\subsection{Behavior in non-bursting model}\label{sec:nonburst}
The low rate of model 3 (c.f.~Fig.~\ref{fig:o15ne19_rate}) stops the bursts and causes a slowly oscillating luminosity as seen in Fig.~\ref{fig:Lt}.
This oscillating behavior will be explained now.

\begin{figure}[t]
\plotone{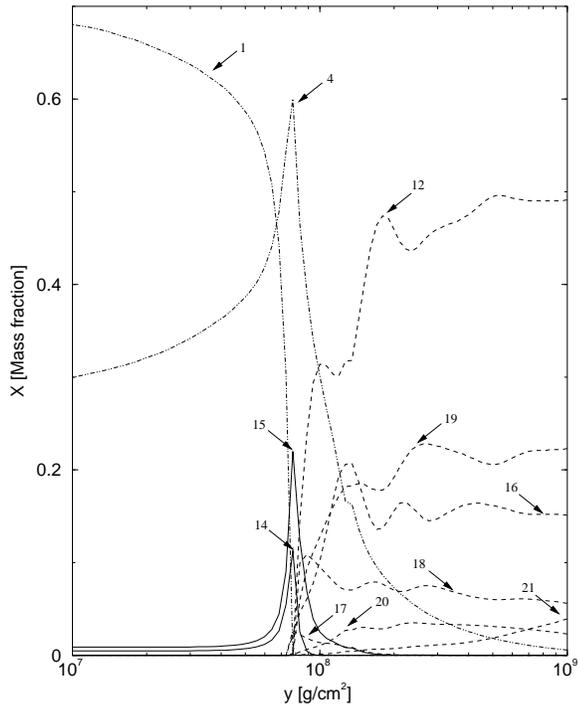}
\caption{The mass fraction of H, ${}^4\textrm{He}$, and the summed mass fractions of A=12, 14 -- 21 nuclei for the stable steady state burning obtained by using the lower estimate of the ${}^{15}\textrm{O}(\alpha,\gamma)$${}^{19}\textrm{Ne}$ reaction rate as a function of column density. This snapshot is taken at a time where the envelope is hot and the helium burning front is at  $y \approx 8\times 10^7$ g cm$^{-2}$, its uppermost level.}\label{fig:lower} 
\end{figure}

A lower ${}^{15}\textrm{O}(\alpha,\gamma)$${}^{19}\textrm{Ne}$ reaction rate restricts the outflow from the HCNO cycle, thereby causing a significant build up of the HCNO mass fraction at an earlier point (see Fig.~\ref{fig:lower}).

The large $X_{HCNO}$ mass fraction in turn converts hydrogen into helium at an increasing rate, fully depleting hydrogen at $y \sim 8 \times 10^7$ g cm$^{-2}$. 
The associated energy generation raises the temperature of the envelope to $T\sim 3\times 10^8$ K and increases the rate of the 3-$\alpha$ reaction, transforming the helium into ${}^{12}\textrm{C}$. 
This is immediately transformed into ${}^{14}\textrm{O}$ and ${}^{15}\textrm{O}$ as long as hydrogen is present. 
The higher temperature moves the helium burning front further out at lower densities, destroying hydrogen through the HCNO cycle up to column densities of $y \sim 6 \times 10^7$ g cm$^{-2}$. 

After the hydrogen has been destroyed, energy generation decreases and the layer cools as it is advected down. Since there is no hydrogen, the 3-$\alpha$ reaction creates ${}^{12}\textrm{C}$, which may capture another $\alpha$ particle to become ${}^{16}\textrm{O}$.

As the temperature of the envelope is now lower, the 3-$\alpha$ reaction increases $X_{HCNO}$ at a slower rate, whence hydrogen is burned slower.
This allows hydrogen to survive to a depth of about $y \sim 1 \times 10^8$ g cm$^{-2}$ 
 until the increasing temperatures finally speed up the 3-$\alpha$ reaction to make sufficient HCNO material to burn hydrogen off again. As a result the burning front of the hydrogen slowly moves back and forth between  $y \sim 6 \times 10^7$ g cm$^{-2}$ and $y \sim 1 \times 10^8$ g cm$^{-2}$. 

Even while burning in this stable mode (as defined by a one-dimensional model with a lower estimate of the ${}^{15}\textrm{O}(\alpha,\gamma)$${}^{19}\textrm{Ne}$ reaction rate) the neutron star may burst due to other reasons which are not considered in this model, e.g., localized burning or changes in the local accretion rate due to accretion instabilities. 
In this case a low ${}^{15}\textrm{O}(\alpha,\gamma)$${}^{19}\textrm{Ne}$ reaction rate will cause a restricted outflow of ${}^{15}\textrm{O}$ during the burst, thus preserving HCNO material for the next burst.
This causes the HCNO cycle to run comparably faster until the subsequent burst, which means that more hydrogen is depleted by stable burning, thereby increasing the observable $\alpha$-parameter of the X-ray burst source. 

\section{Implications for superbursts}\label{sec:superburst}
Superbursts have been observed from a number of sources \citep{Wijnands01,Kuulkers02b,Cornelisse02} and have been theorized to be caused by carbon burning in the neutron star ocean \citep{Cumming01}. 
The narrow parameter space in which superbursts are possible is given in \cite{Cumming01}, but current calculations \citep{Schatz99,Schatz01,Woosley04,Fisker04} have not been able to generate the required amounts of ${}^{12}\textrm{C}$; in addition, these calculations leave residual ${}^{4}\textrm{He}$ which will stably burn the ${}^{12}\textrm{C}$ into ${}^{16}\textrm{O}$ as matter sinks deeper into the ocean.

However, Fig.~\ref{fig:lower} shows that the ashes of stable burning almost solely comprise ${}^{12}\textrm{C}$ along with other isotopes, mainly ${}^{16}\textrm{O}$, ${}^{19}\textrm{F}$, ${}^{18}\textrm{F}$ and ${}^{18}\textrm{O}$.
This means that stable burning of this kind need not be maintained at all times, but can work for short periods of a few months, which are sufficient to generate the required carbon for the next superburst.
\cite{Cumming04a} estimate a duration of weeks for the cooling time of a superburst, during which the ocean is still hot enough to prevent X-ray bursts from occuring. 
This means that continuously accreting matter burns stably during this time, whence a significant fraction of the nuclei shown in Fig.~\ref{fig:lower} could be generated.

One might also speculate that the local accretion rate is not the same over the entire star, and that high local accretion rates might cause material to burn stably.
It is therefore important to establish the value of the critical accretion rate where unstable burning turns stable (see \cite{Rembges99} and \cite{Fisker03}) as well as obtain a good model of how the accretion flow distributes itself on the star.

\section{Conclusions}\label{sec:conclusion}
We have demonstrated the significance of the ${}^{15}\textrm{O}(\alpha,\gamma){}^{19}\textrm{Ne}$ reaction to X-ray burst behavior and suggested a lower limit on the rate which could and should be pursued experimentally.

Long intermittent periods of stable burning, which may last for months, have been observed from burst sources; a lower value of the ${}^{15}\textrm{O}(\alpha,\gamma){}^{19}\textrm{Ne}$ reaction rate will help stabilize stable burning. 
The reaction rate determines the value of the critical accretion rate. 

In addition we have shown a way to generate the necessary carbon for the superburst theory to be valid. We have also shown that several different kinds of $A\lesssim 20$ nuclei are generated; applying this to superburst calculations requires proper knowledge of the fusion rates between these nuclei.

\acknowledgements
We thank H.~Schatz and S.~Austin for suggesting an investigation of this rate and A.N.~Ostrowski for discussions. 
JLF, JG, and MW were supported by NSF-PFC grant PHY02-16783 through the Joint Institute for Nuclear Astrophysics\footnote{see \email{http://www.JINAweb.org}}. BD acknowledges support from the Natural Sciences and Engineering Research Council of Canada.

\end{document}